\documentclass[conference]{IEEEtran}
\IEEEoverridecommandlockouts
\usepackage{cite}
\usepackage{amsmath,amssymb,amsfonts}
\usepackage{algorithmic}
\usepackage{graphicx}
\usepackage{textcomp}
\usepackage[table, xcdraw]{xcolor} 
\usepackage{hyperref}
\usepackage{url}
\usepackage{array}  
\usepackage{booktabs} 

\usepackage{colortbl}      

\usepackage{soul}  
\usepackage{booktabs}		

\usepackage{comment,svg,subfigure}
\newcommand{\cscCompact}{\mathrm{C\mkern-1mu S\mkern-1mu C}}
\newcommand{\surplusCompact}{\mathrm{s\mkern-1mu u\mkern-1mu r\mkern-1mu p\mkern-1mu l\mkern-1mu u\mkern-1mu s}}
\newcount\Comments  
\newcommand{\kibitz}[2]{\ifnum\Comments=1{\color{#1}{#2}}\fi}

\def\BibTeX{{\rm B\kern-.05em{\sc i\kern-.025em b}\kern-.08em
    T\kern-.1667em\lower.7ex\hbox{E}\kern-.125emX}}
\begin{document}

\title{Fairness of Energy Distribution Mechanisms in Collective Self-Consumption Schemes \\

}
\author{\IEEEauthorblockN{Benoit Couraud$^{1}$, Valentin Robu$^{2}$, Sonam Norbu$^{1}$, Merlinda Andoni$^{1}$,  Yann Rozier$^3$,  Si Chen$^{1}$, Erwin Franquet$^3$, \\ Pierre-Jean Barre$^3$, Satria Putra Kanugrahan$^{1}$, Benjamin Berthou$^4$, David Flynn$^{1}$}
	
	\IEEEauthorblockA{$^{1}$James Watt School of Engineering, University of Glasgow, Glasgow (UK),
        $^2$Intelligent and Autonomous Systems Group,\\ CWI, Amsterdam (The Netherlands),
        $^3$Université Côte d’Azur, IMREDD \& Polytech'Lab, France, 
        $^4$Enogrid, France
        }
}  
\maketitle

\begin{abstract}
In several European countries, regulatory frameworks now allow households to form energy communities and trade energy locally via local energy markets (LEMs). While multiple mechanisms exist to allocate locally produced energy among members, their fairness  remains insufficiently understood—despite energy justice being a key concern for communities.
This paper first provides a thorough description of the collective self-consumption (CSC) process in France, offering a real-world framework for researchers. We then review the main types of fairness relevant to LEMs and identify appropriate indicators for each, including a new scalable indicator to evaluate meritocratic fairness.
Using simulations across 250 randomly generated residential communities of 20 households, we assess and compare fairness across different LEM distribution mechanisms. Results show that average financial savings reach 12\% with 40\% PV uptake.
Among the four widely used LEM mechanisms assessed, glass-filling with prioritization yields the highest egalitarian and min-max fairness. Double auction and pro rata schemes promote meritocracy, while standard glass-filling offers a strong balance across fairness objectives.


\end{abstract}


\section{Introduction}
\label{sec:intro}
    
The energy transition requires the deployment of distributed energy resources, such as solar photovoltaic (PV) systems. However, recent tariff changes in many countries have further reduced the financial compensation for exporting residential solar PV electricity to the grid~\cite{cre2024photovoltaïque}, thereby weakening the incentive to invest in such systems. In France, for example, feed-in-tariffs have been divided by 3 for residential PV in 2025~\cite{cre2024photovoltaïque}. In this context, local energy communities and collective self-consumption (CSC) schemes are emerging as attractive alternatives to support solar PV investment. CSC refers to local initiatives in which producers and prosumers (consumers with local generation) can share or sell surplus energy with nearby community members.
In response to the European legal framework 
which defines Renewable Energy Communities (RECs) and Citizen Energy Communities (CECs) as legal entities enabling citizens, small businesses, and local authorities to collectively generate, consume, store, and trade energy~\cite{EUdirective2019-944}, France has established a national framework to enable CSC initiatives~\cite{codeenergieL315-2}. Since 2021, 883 CSC-based local energy communities have been created, each involving on average 10 consumers and 2 producers~\cite{enedisACC2024}. These communities offer financial advantages: producers can earn more by selling electricity directly to neighbors than through feed-in tariffs, while consumers benefit from access to local green electricity at prices lower than those offered by traditional suppliers.

Still, the establishment of CSC energy communities requires members to select an energy distribution mechanism or local energy market (LEM) to determine who benefits from local energy exchanges and at what price. Several  of these  mechanisms are currently offered by major energy community stakeholders, ranging from static or pro-rata allocation methods provided by the French Distribution System Operator (DSO)~\cite{enedis_autoconsommation_guide}, to more advanced approaches that aim to distribute benefits equally among all participants~\cite{enogrid_cle_dynamique_simple}. Since energy communities are intended to deliver economic and social benefits~\cite{EUdirective2019-944}, fairness has become a central concern. Most CSC initiatives seek to implement LEM that promote energy justice and ensure fair outcomes for all participants. However, communities often lack guidance on which approach best aligns with their values — such as equity, meritocracy, or need-based fairness, which are three different approaches of fairness.
As a result, fairness in energy communities has become an active area of research, with studies focusing on the evaluation of fairness indicators~\cite{DYNGE2025125463} or on the development of novel LEM mechanisms aiming for greater fairness~\cite{GJORGIEVSKI2022124246}. Fairness of transactive energy benefits distribution has also been addressed in the broader context of local energy systems~\cite{SOARES2024123933}. However, none of these studies actually compare the fairness of the main energy distribution mechanisms proposed to real-world energy communities. Furthermore,  many of these studies either fail to comprehensively address the full spectrum of fairness types — such as utilitarian~\cite{SOARES2024123933}, or meritocratic fairness~\cite{DYNGE2025125463}. Finally, research works often lack integration with actual tariff structures from countries that have already implemented CSC regulatory frameworks. 
\begin{figure*}[ht!]
\vspace{-3mm}
	\centerline{\includegraphics[width=0.95\textwidth]{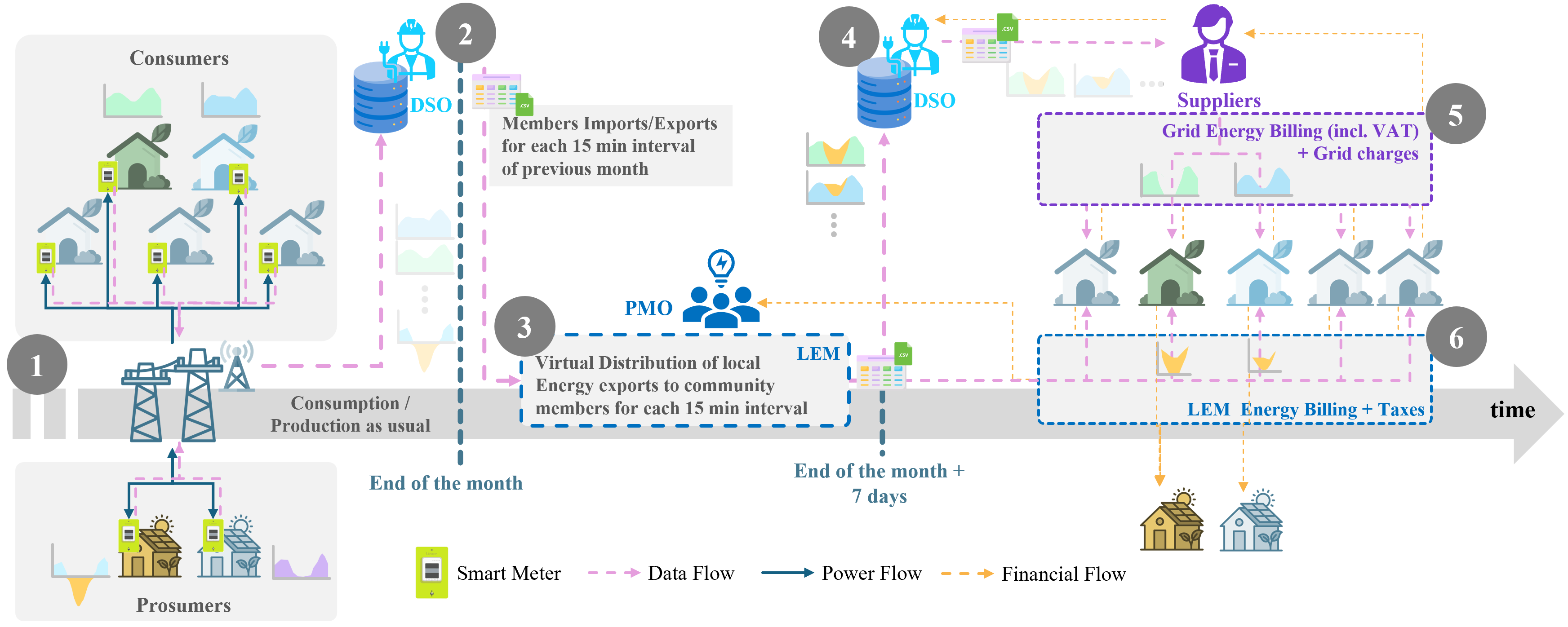}}
	\caption{Timeline and workflow of a standard CSC implementation in France.}
    \vspace{-3mm}
	\label{fig:CSC_Flow}
\end{figure*}
To address these gaps, we first aim to harmonize the various perspectives on fairness in LEM and introduce a new scalable fairness indicator tailored to meritocracy. We then conduct simulations on 250 randomly generated residential energy communities with real households data over one year under varying levels of solar PV adoption to evaluate four major energy distribution mechanisms currently proposed to CSC operations in France. These mechanisms are assessed against the main fairness types using the most appropriate fairness indicators.

The remainder of this paper is structured as follows: Section~\ref{sec:context} outlines the financial and operational workflows of CSC schemes in France. Section~\ref{sec:fairness} describes the distribution mechanisms analyzed, along with the different fairness types and their indicators. Finally, Section~\ref{sec:experiment} compares the fairness of the selected LEM mechanisms.

\section{Context of CSC in France}
\label{sec:context}
This section provides a detailed explanation of how CSC and LEM are implemented in certain European countries (e.g., France and Italy), offering real-world context that researchers can build upon in future work. Indeed, since 2021, a legal framework in France has required DSOs to support the formation of CSC schemes~\cite{codeenergieL315-2}. A CSC involves establishing a legal entity, the Personne Morale Organisatrice (PMO), which coordinates the community and registers its members with the DSO as consumers, producers, or prosumers, ensuring injection rights for the latter two.
Typically, producers and prosumers also have contracts with EDF OA (a French public utility), which purchases surplus electricity at a fixed feed-in tariff valid for 20 years. However, this rate has dropped significantly, from 12.69 c€/kWh (excl. VAT) for installations registered in 2024 to just 4 c€/kWh in 2025~\cite{cre2024photovoltaïque}.
As illustrated in Fig.\ref{fig:CSC_Flow}, the CSC process involves several steps. First, members produce and consume energy as usual, although some communities may coordinate flexibility to optimize self-consumption (Step 1). At the end of each month (Step 2), the DSO provides the PMO with 15-minute metering data for all members. Then (Step 3), the PMO runs an a posteriori Local Energy Market (LEM), computed based on ex-post consumption and production data, to allocate exports from producers/prosumers to consumer imports for each time step. The LEM also sets local energy prices, which may be uniform, fixed or time-/pair-specific. Two configurations are permitted: Peer-to-Community (P2C), where all trades occur between members and the community as a whole; and full Peer-to-Peer (P2P), where trades occur directly between individuals\cite{SOUSA2019367}.
To ensure mutual benefit, local prices must remain below supplier retail prices (for consumers) and above feed-in tariffs (for producers), as discussed later. In Step 4, the PMO sends allocation data to the DSO, specifying energy transactions by time step and participant (per the P2P or P2C model). The DSO updates metering records accordingly, allowing suppliers to correctly bill for grid imports/exports and account for CSC volumes.
Finally (Step 5), suppliers bill consumers and producers for net grid usage, including applicable taxes and network charges for both grid imports and community self-consumption. The consumer can chose wether these network charges are of the same amount for both, or lower (higher) for CSC (grid) imports respectively. Separately, community members are invoiced for local trades, either directly or via an aggregator, based on the agreed LEM price. A share of the revenue may also go to the PMO to cover its costs. 

\begin{figure}[ht]
\centering
\includegraphics[width=0.85\columnwidth]{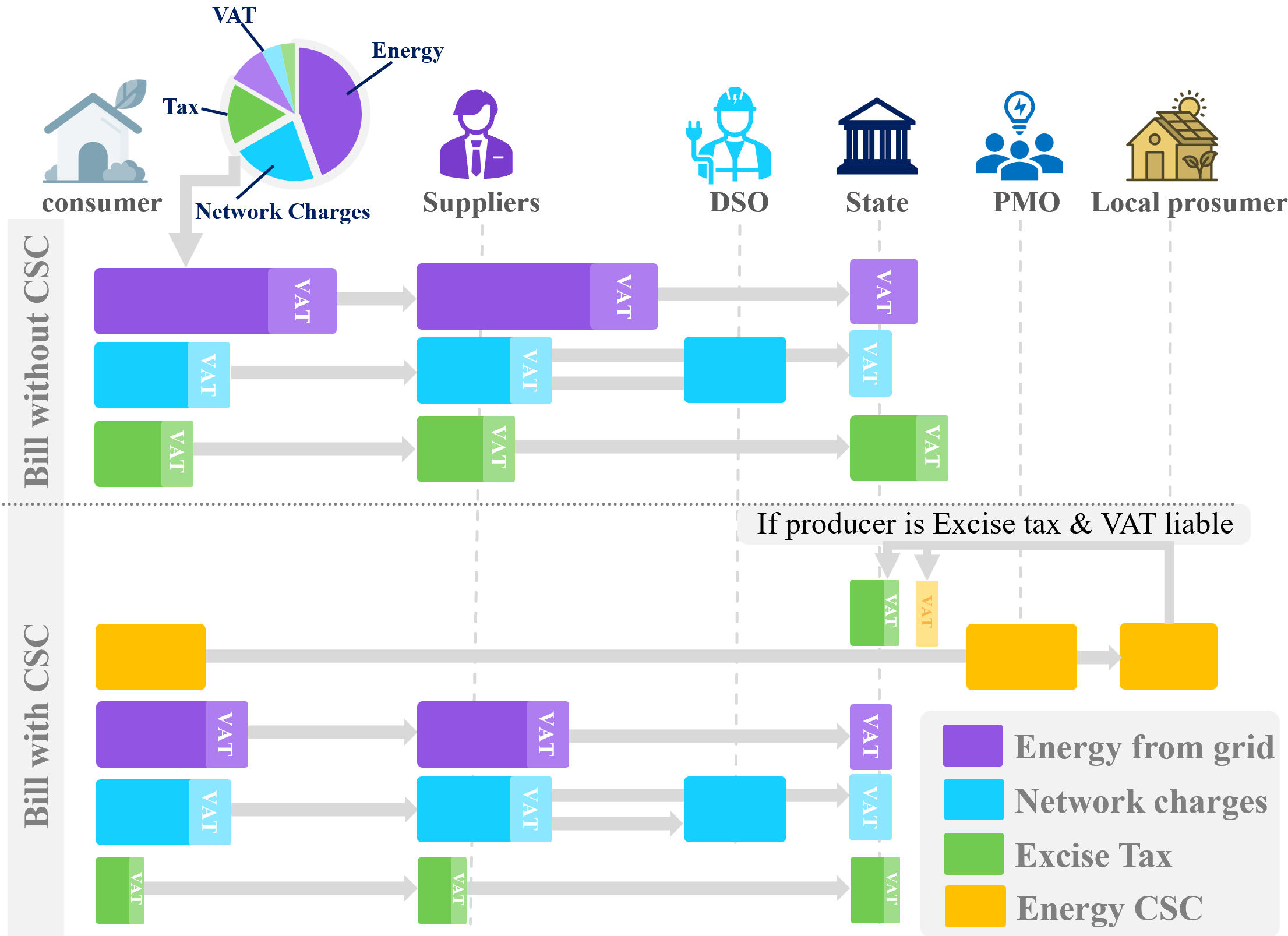}
\caption{Breakdown of consumers bills and financial flows (gray arrows) between the main stakeholders with and without CSC.}
\label{fig:Bill_Breakdown}
    \vspace{-3mm}
\end{figure}

To simulate correctly the benefits of CSC operations, it is necessary to model correctly the pricing mechanisms in CSC schemes. To this end, the breakdown of the variable part of an electricity bill with and without CSC is shown in Fig.\ref{fig:Bill_Breakdown}. 
In a standard electricity bill without CSC, the consumer pays the supplier for the energy provided (including VAT), an indirect tax for consumers named "excise tax" of €29.98/MWh for residential consumers (lower for others types)~\cite{DGFiP2025Accises}, and network charges—both subject to VAT. The supplier then transfers the network charges to the DSO and the taxes to the state.

In a CSC operation, assuming the same amount of energy is consumed at the same times as in the previous paragraph, but partly supplied by local prosumers, the network charges remain unchanged and are still paid to the supplier. The excise tax paid to the supplier only applies to the portion of electricity drawn from the grid. Consumer 
$i$ then pays either the community (in a P2C configuration) or the prosumers  (in a full P2P scheme) an amount equal to $\sum_t E_{ij}^{\cscCompact, t}\cdot \pi_{ij}^t$ where $E_{ij}^{\cscCompact, t}$ is the energy self-consumed by consumer $i$ at time step $t$ and produced by producer $j$, and $\pi_{ij}^t$ is the LEM price at time $t$ between consumer $i$ and producer $j$. This price may be uniform across all members.
Depending on the status of the producer, he may have to deduct from these gains $\left(\sum_t E_{ij}^{\cscCompact, t}\cdot \pi_{ij}^t\right)$ the cost of excise duties (if his installed capacity exceeds 1 MW), which vary based on the type of consumer $i$, and VAT if he is VAT-liable (i.e., companies with sufficiently high turnover). As a result, for producer $j$, the gain from selling energy within the community—compared to standard export tariffs—corresponds to the difference between the export tariff (excl. VAT by default) and the payment received from consumers $\left(\sum_i\sum_t E_{ij}^{\cscCompact, t}\cdot \pi_{ij}^t\right)$ reduced from VAT (if applicable), and from excise duties (if applicable). For consumers, the financial benefit of joining a CSC lies in the difference between the cost of locally sourced energy and the cost of purchasing the same quantity from the supplier—excluding network charges only.

\section{Fairness of Energy Distribution Mechanisms}
\label{sec:fairness}
The aim of this work is to assess the fairness of several approaches available to energy communities to virtually distribute local energy among consumers/producers (step 3 of Fig. \ref{fig:CSC_Flow}). We first describe the energy distribution algorithms considered in this work, and then present how to assess their fairness in terms of financial benefits.
\subsection{Energy distribution mechanisms}
\label{sec:distributionAlgorithms}
Several standard energy distribution approaches are proposed by some companies that manage French CSC operations~\cite{enogrid_cle_dynamique_simple}, and will be assessed in this work.  

The first mechanism corresponds to the \textit{pro-Rata} approach, that is proposed by default by the DSO, which consists in the allocation of the community energy proportionally to the distribution of energy imports or exports within the community, as shown in Eq. \ref{eq:proRata}.

\begin{equation}
\begin{aligned}
\label{eq:proRata}
E^{\cscCompact, t}_{i} = \frac{E^{t}_{i}}{\sum_{k\in \mathcal{C}_t}E^{t}_{k}}\cdot E^{\cscCompact, t} \\
P^{\cscCompact, t}_{j} = \frac{P^{t}_{j}}{\sum_{k\in \mathcal{P}_t}P^{t}_{k}}\cdot E^{\cscCompact, t} 
\end{aligned}
\end{equation}

\noindent where $E^{\cscCompact, t}_{i}$ and $P^{\cscCompact, t}_{j}$ are the quantities of locally allocated energy imported by consumer 
$i$ and exported by producer $j$ at time $t$, respectively. $E^{t}_{i}$ and $P^{t}_{j}$ denote the total energy imported/exported by consumer $i$ and producer $j$ at time $t$, as reported by the DSO to the PMO. The sets $\mathcal{C}_t$ and $\mathcal{P}_t$ represent the consumers and producers who have imported or exported energy at time step $t$, respectively. The total amount of energy available for allocation within the LEM at time $t$ is given by: 
$E^{\cscCompact, t} = \min\left(\sum_{k\in \mathcal{C}_t}E^{t}_{k}, \sum_{k\in \mathcal{P}_t}P^{t}_{i}\right)$. If, at time step $t$, a producer is awarded to sell all his production to the community, then $P^{\cscCompact, t}_{j} = P^{t}_{j}$. The pro-rata will allocate more local energy to consumers who consume the most, which does not correspond to equal fairness.

The second distribution mechanism assessed is the \textit{glass-filling} approach, so named because it follows the logic of several interconnected glasses being filled. This is implemented through an iterative process, described below: At time $t$ and iteration $0$, each consumer is initially allocated an equal share of energy, defined as
$E^{\cscCompact^0, t}_{i} = \min\left(E^{t}_{i},\frac{E^{\cscCompact, t} }{\left|\mathcal{C}_t^{0,^*}\right|}\right)$, where ${C}_t^{k,^*}$  denotes the set of consumers at time $t$ and iteration $k$ who are still eligible to receive additional community energy in the next iteration, that is, those for which $E^{\cscCompact^{k-1}, t}_{i} < E^{t}_{i}$. It is initialized as ${C}_t$.

Then, at iteration $k$, the energy allocated to consumer $i$ is: 
\begin{equation}
\begin{aligned}
\label{eq:glassFilling}
E^{\cscCompact^k, t}_{i} = \min\left(E^{t}_{i},E^{\cscCompact^{k-1}, t}_{i}+\frac{P^{\surplusCompact^{k},t}}{\left|\mathcal{C}_{t}^{k,^*}\right|}\right)\\
P^{\surplusCompact^k,t} = E^{\cscCompact, t}  - \sum_{i \in \mathcal{C}_t}E^{\cscCompact^{k-1}, t}_{i}
\end{aligned}
\end{equation}

A similar definition applies to allocate $E^{\cscCompact, t}$ among producers. In this approach, at each time step $t$, all members initially receive an equal share of energy, and surpluses from low-consumption households are iteratively redistributed to higher-consuming ones.

The third distribution mechanism, \textit{Prioritized glass-filling}, builds upon the second, but adds a fairness criterion: at each time step $t$, priority is given to consumers (or producers) who have received the least amount of local energy since the beginning of a rolling year. At each time step $t$, consumers (or producers) are ranked by their cumulative allocated energy up to that time $t$. Multiple members can share the same priority level. For each priority level $l$, starting from the highest, $E^{\cscCompact, t}$ is distributed using the same glass-filling approach (with iteration index $k$ restarting at $0$ for each priority level). In this case, $\mathcal{C}_{t}^{k,^*}$ is replaced by $\mathcal{C}_{l,t}^{k,^*}$, which represents the set of consumers (or producers) with priority level $l$ who can still receive community energy at iteration $k$.

Finally, in the fourth distribution mechanism assessed, we implement a uniform-price \textit{double auction} to clear the LEM. In this mechanism, at each time step $t$ all demand bids (consumers) and production offers (producers) are submitted simultaneously for each time step, and consist of the quantity of energy consumed $E^{t}_{i}$ and produced $P^{t}_{j}$ respectively, along with the supplier's energy price (incl. taxes (VAT and excise tax), but excl. network charges) for consumers, and the supplier's export tariff price for the producers (incl. VAT and maximum applicable excise tax in the community, if they apply to the producer). 
The CSC energy $E^{\cscCompact, t}$ is then distributed using the third approach with a priority on price this time (sorted in descending order for consumers, and ascending order for producers prices), instead of previous energy allocated.

For all allocation mechanisms, the LEM price $\pi_t$ is considered uniform across all consumers and producers. It is computed as the average between (i) the highest export price among the producers awarded local energy—defined as their standard export tariff price including VAT and the maximum applicable excise tax if they apply to the producer— and (ii) the lowest import price among the consumers receiving local energy—defined as the consumer's supplier’s energy price including VAT and excise tax, but excluding network charges as it is already paid to the consumer's supplier independently.

\subsection{Fairness assessment in LEM}
Fairness can be defined in various ways, depending on the specific fairness objective to be achieved. Equality-based fairness in a LEM aims for an equal distribution of locally produced energy among community members, regardless of their individual profiles. It can be assessed using indicators such as the Jain index, Gini coefficient, or Quality of Service (QoS). Among these, the Jain index is the most commonly used, as it quantifies the equality of distribution and yields a value of 1 for perfectly equal outcomes. This is formalized in Table~\ref{tab:fairnessType}, where  $u_i = B_i - \hat{B}_i$ represents the financial benefit gained by community member $i$ from participating in the LEM, with $B_i$  the annual bill without LEM participation, and $\hat{B}_i$  the bill with LEM participation. 
 
Then, building on John Rawls’ maximin principle, min-max fairness aims to maximize the minimum utility received by any participant. Rooted in Rawls’ theory of justice, this approach holds that social and economic arrangements should be designed to benefit the least advantaged members of society. In the context of LEMs, this principle helps ensure that the participant receiving the lowest benefit is prioritized, thereby protecting the most vulnerable. The Min-Max Ratio is typically used to assess this form of fairness.

Meritocratic fairness is another important fairness concept that aims to reward community members in proportion to the value they contribute to the community welfare. While this value can encompass social, environmental, or technical contributions, most studies on LEM implementations and simulations limit the assessment to financial contributions, due to a lack of data on non-financial impacts.
Meritocratic fairness is commonly assessed using the Shapley value, which allocates rewards based on each member’s marginal contribution to the collective outcome. However, Shapley value computations become intractable as the number of participants increases, due to their combinatorial complexity. Although some approaches such as "last marginal contribution" can efficiently approximate the Shapley value, they still require to run LEM simulation for $N$ new communities \cite{CREMERS2023120328}. To address this limitation, we propose an alternative indicator that evaluates each member $i$’s contribution to the energy trades of the rest of the community. As detailed in Table~\ref{tab:fairnessType}, this indicator is based on the difference between a member’s actual savings from joining the LEM, denoted by $u_i$, and their ideal meritocratic share $\hat{u}_i$. The latter is calculated as the product of the member’s normalized contribution ($\frac{C_i}{\sum_j C_j}$), and the total community savings ($\sum_j u_j$). Here, $C_i$  represents member $i$’s contribution to the community trades, defined as:
\begin{equation}
\label{eq:contribution}
C_i = \sum_t \left( -\operatorname{sign}(\hat{E}_i^t)\cdot\min\left(|\hat{E}_i^t|,|\hat{E}_{-i}^t|\right)\cdot \hat{E}_{-i}^t\right)
\end{equation}

\noindent where $\hat{E}_i^t$ is the net energy injection of member $i$ at time $t$—positive for export, and negative for import—and  $\hat{E}_{-i}^t$ is the sum of net energy injections from all other members at the same time. Consequently, $C_i$
  will be high for members who export energy when others are importing, or import when others are exporting, as they contribute more to the overall benefit of the community. The use of the $\min$ function prevents an unfair advantage for members who overproduce or overconsume.
  
Beyond these three main approaches of fairness, other types of fairness have also been proposed in the literature. Utilitarian fairness aims to maximize the total welfare of the community by summing all individual gains, regardless of members' specific gains, needs or contributions. Need-based fairness assigns weights to members’ gains according to their specific  socio-economic needs. These weights are typically derived from exogenous data (e.g., income level, energy vulnerability), which is often unavailable—limiting the analysis of this fairness type in many studies, including this work that will not address it for the same reason.
%

\begin{table}[ht]
        \vspace{-1mm}
    \caption{Types of Fairness and Corresponding Indicators in Local Energy Distribution.}
    \label{tab:fairnessType}
    \small
    \centering
    \begin{tabular}{>{\raggedright\arraybackslash}m{2.3cm} 
                    >{\raggedright\arraybackslash}m{2.3cm} 
                    >{\raggedright\arraybackslash}m{3.4cm}}
        \toprule[\heavyrulewidth]\toprule[\heavyrulewidth]
        \textbf{Fairness Type } & \textbf{Indicator} & \textbf{Formula} \\
        \midrule
    \arrayrulecolor[gray]{0.8} 

        Equality  & Jain index & 
        $\frac{(\sum_{i=1}^{n} u_i)^2}{n \cdot \sum_{i=1}^{n} u_i^2}$ \\

        \specialrule{0.01pt}{1pt}{1pt}
  
        Min-Max  & Min-Max ratio & 
        $\frac{\min_i u_i}{\max_i u_i}$ \\
        \specialrule{0.01pt}{1pt}{1pt}


    Meritocratic      & Meritocratic index  & 
    $\sqrt{\frac{1}{n} \sum_i (u_i - \hat{u}_i)^2}$, where $\hat{u}_i = \frac{C_i}{\sum_j C_j} \sum_j u_j$ \\
        \specialrule{0.01pt}{1pt}{1pt}

        Utilitarian  & Social welfare & 
        $\sum_{i=1}^{n} u_i$ \\
        \specialrule{0.01pt}{1pt}{1pt}


        Need-based  & Weighted utility  & 
        $\sum_{i=1}^{n} w_i \cdot u_i$, where $w_i$ reflects user needs\\


             \arrayrulecolor{black} 

        \bottomrule[\heavyrulewidth]
    \end{tabular}
        \vspace{-3mm}
\end{table}

\section{Experimental Results}
\label{sec:experiment}
To evaluate the fairness of the 4 energy distribution algorithms described in Section~\ref{sec:distributionAlgorithms}, we conducted annual simulations for 50 energy communities, each comprising 20 members randomly assigned either a fixed or a time-of-use tariff. The consumption profiles of these members were different for each community, randomly selected from a dataset of 200 consumers provided by the Thames Valley Vision project. We simulated the outcomes of the 4 LEM schemes across 5 different levels of PV uptake (yielding to 250 communities overall), ranging from 0 to 80\%, assuming residential PV assets of  3~kW, with 30\% of PV newly installed, i.e. with reduced export tariff (4~c€/kWh). Solar production data has been collected from a  solar plant from south east of France.

\begin{figure}[ht]
\centering
\vspace{-2mm}
\includegraphics[width=0.9\columnwidth]{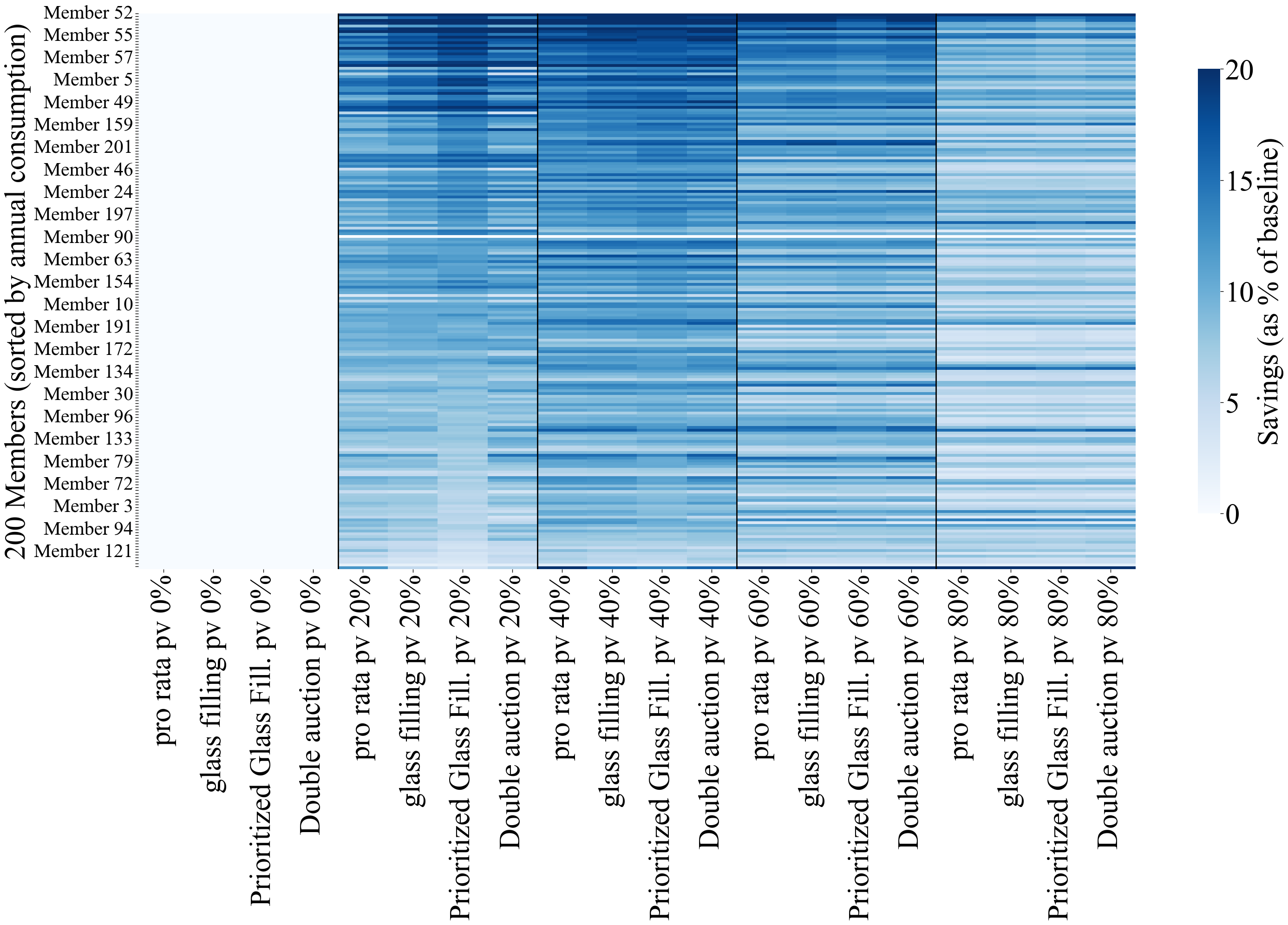}
\caption{Percentage of annual bill savings averaged over all simulations, with the 200 members ordered by annual energy consumption (descending order).}
\vspace{-2mm}
\label{fig:Bill_savings}
\end{figure}

Figure~\ref{fig:Bill_savings} presents the average percentage of bill reduction across all 200 consumers, plotted by LEM scheme and PV uptake level. Members are ordered from the lowest (top) to highest annual energy consumption (bottom). The results show that in average, LEM without flexibility can lead to a percentage annual bill reduction of 12\% for solar PV uptake of 40\%.
Beyond 40\% PV uptake, the benefits tend to decrease in percentage due to the decreased trading potential when most members already have PV. As the figure illustrates percentage savings, lower-consumption users appear to benefit more in relative terms, which would be reversed if absolute savings were considered. Moreover, the distribution of savings varies by LEM scheme. For example, up to 40\% PV uptake, the prioritized glass-filling mechanism offers higher percentage savings to low-consumption users, while high-consumption users benefit less. Beyond this threshold, the differences between distribution schemes become less pronounced.

\vspace{-1mm}

\begin{figure}[ht]
\vspace{-2mm}
\centering
\includegraphics[width=0.95\columnwidth]{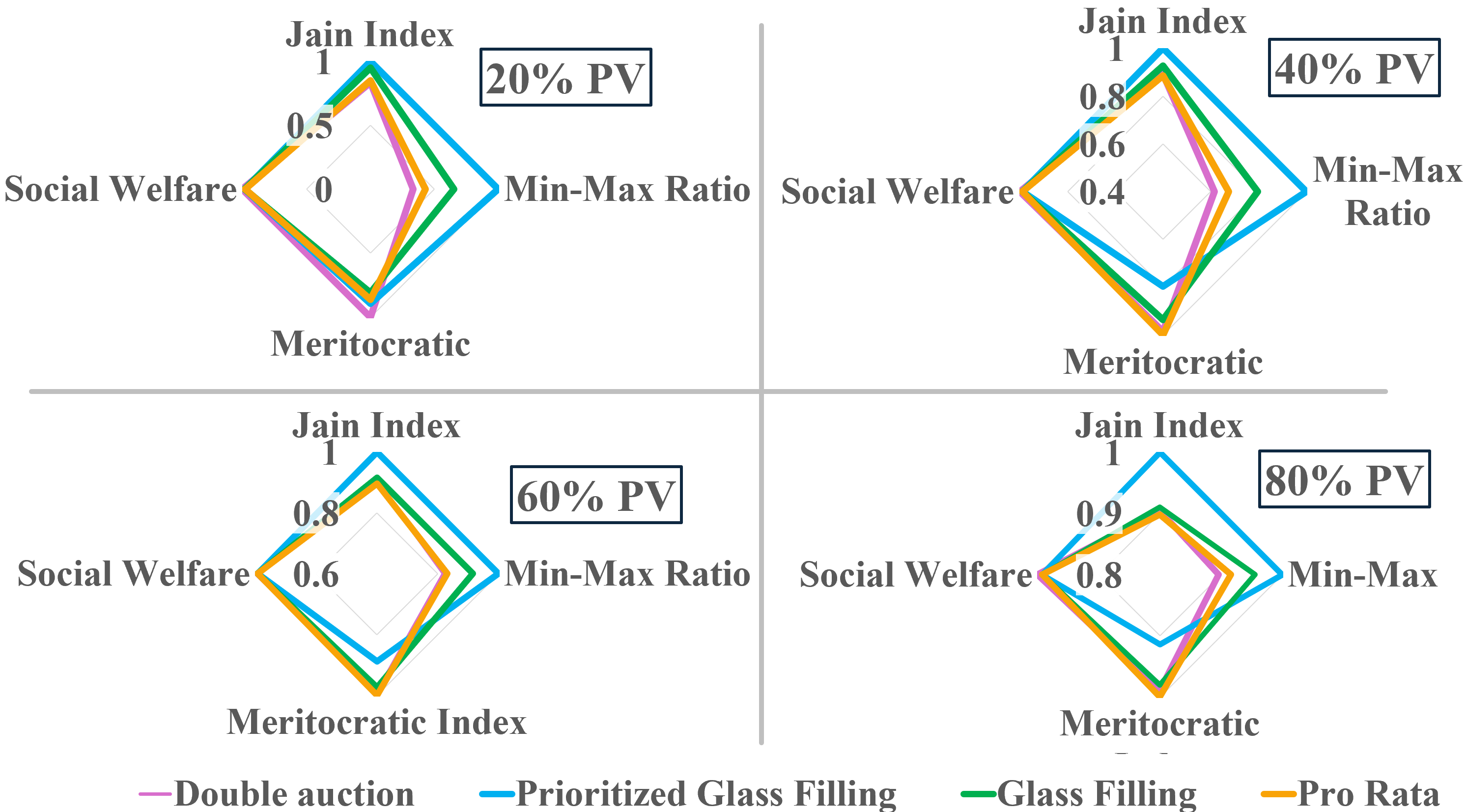}
\caption{Comparison of fairness of distribution mechanism under different solar PV penetration rates.}
\label{fig:Social_Metrics}
\vspace{-3mm}
\end{figure}

These observations are further supported by the normalized fairness indicators shown in Figure~\ref{fig:Social_Metrics} across varying PV uptake levels. As expected, the prioritized glass-filling approach yields higher Jain index and Min-Max ratio values, indicating stronger egalitarian and equalitarian fairness. In contrast, the double auction and pro-rata mechanisms perform better in terms of meritocratic fairness, while utilitarian fairness is similarly achieved across all mechanisms, as it considers only the community as a whole. Finally, although there is not one distribution approach that addresses all types of fairness, the simple glass-filling method consistently offers a compromise between meritocratic and egalitarian fairness principles.

\section{Conclusion}
\label{sec:conclusion}
Using a real-case LEM configuration with a detailed pricing model, this paper compares the fairness of various energy distribution algorithms. Simulations on 250 randomly generated communities show that average financial savings for residential members peak at 12\% with 40\% PV uptake. Among the algorithms analyzed, the prioritized glass-filling approach—which prioritizes members who received the least local energy over the year—performs best in terms of egalitarian and equalitarian fairness. In contrast, double auction and pro-rata methods promote meritocratic fairness, rewarding members proportionally to their contribution, as captured by a novel scalable fairness indicator. As a middle ground, the standard glass-filling method offers a compromise between egalitarian and meritocratic goals. These findings can help energy communities choose a distribution mechanism that aligns with their interpretation of fairness. However, they should be complemented by community engagement efforts to co-develop new algorithms that reflect local priorities and values, thereby enhancing social acceptability.

\section*{Acknowledgment}
This work was supported  by EPSRC projects HI-ACT (EP/X038823/2), and DISPATCH [EP/V042955/1].

\bibliographystyle{ieeetr}
\bibliography{main}

\end{document}